AI-Supported Platform for System Monitoring and Decision-Making in
Nuclear Waste Management with Large Language Models–25367

Dongjune Chang[1], Sola Kim[2], and Young Soo Park[3]
[1]School for Engineering of Matter, Transport and Energy, Arizona State University
[2]School of Sustainability, Arizona State University
[3]Argonne National Laboratory

ABSTRACT

Nuclear waste management requires rigorous regulatory compliance assessment, demanding advanced decision-support systems capable of addressing complex legal, environmental, and safety considerations. This paper presents a multi-agent Retrieval-Augmented Generation (RAG) system that integrates large language models (LLMs) with document retrieval mechanisms to enhance decision accuracy through structured agent collaboration. Through a structured 10-round discussion model, agents collaborate to assess regulatory compliance and safety requirements while maintaining document-grounded responses. Implemented on consumer-grade hardware, the system leverages Llama 3.2 and mxbai-embed-large-v1 embeddings for efficient retrieval and semantic representation. A case study of a proposed temporary nuclear waste storage site near Winslow, Arizona, demonstrates the framework's effectiveness. Results show the Regulatory Agent achieves consistently higher relevance scores in maintaining alignment with legal frameworks, while the Safety Agent effectively manages complex risk assessments requiring multifaceted analysis. The system demonstrates progressive improvement in agreement rates between agents across discussion rounds while semantic drift decreases, indicating enhanced decision-making consistency and response coherence. The system ensures regulatory decisions remain factually grounded, dynamically adapting to evolving regulatory frameworks through real-time document retrieval. By balancing automated assessment with human oversight, this framework offers a scalable and transparent approach to regulatory governance. Future research will explore multi-modal data integration and reinforcement learning to enhance response coherence and decision efficiency. These findings underscore the potential of AI-driven, multi-agent systems in advancing evidence-based, accountable, and adaptive decision-making for high-stakes environmental management scenarios.

INTRODUCTION

Nuclear waste management is a quintessential example of a challenge demanding multifaceted institutional structures, given its blend of technical intricacies, societal concerns, and political dimensions. Frequently referred to as a "wicked problem," this issue defies simple resolutions due to the involvement of varied and often conflicting stakeholders, the need for coordinated actions across different governance levels, and the far-reaching, irreversible nature of the decisions involved [1], [2]. Decision-making here is profoundly shaped by the inherent uncertainties and limitations of data, further complicated by human cognitive constraints that can obstruct effective governance. Ambiguities in technical knowledge about the long-term effects of storage and the unpredictability of human behaviors amplify the difficulty in ensuring safety across extended timescales. Consequently, governance approaches must transcend isolated decisions, incorporating institutional perspectives that address organizational structures and group dynamics. Reconciliation of competing interests and the integration of factors beyond technical considerations are critical, enabling choices to be made amidst persistent uncertainty. The absence of definitive frameworks demands that decision-makers confront the weight of responsibility and long-term implications without a clear resolution [3].

Given the unattainable nature of complete information [4], a focus on reducing uncertainty through an exhaustive analysis of interconnected technical and societal systems becomes imperative. This requires a commitment to avoiding logical errors, cognitive biases, and groupthink [5], where the push for unanimity may suppress critical evaluation and realistic alternatives. By breaking down complex problems into smaller, manageable components, targeted solutions can be crafted and seamlessly integrated into broader strategies. Achieving success in this domain hinges on establishing clear, prioritized objectives that guide meaningful trade-offs while embracing the inherent constraints of forecasting over extended horizons.

AI systems grapple with fundamental difficulties in addressing environmental uncertainties and the ramifications of unforeseen actions. Cross-domain knowledge transfer and adaptive problem-solving offer pathways to manage such uncertainties, yet they underscore the persistent nature of ambiguity in tackling ill-structured problems [6]. The ability to contend with unpredictable outcomes remains a pivotal aspect of AI development, particularly as these systems operate within complex, evolving environments [7], [8]. A central objective lies in equipping AI with the capacity to adjust its problem representations and operational strategies in response to new information [4]. Progress in this area builds on established methodologies, including planning, semantic networks, and dynamic feedback, while extending these approaches to meet the demands of increasingly intricate problem domains. Such systematic advancements pave the way for AI systems to maintain effectiveness amid the fluctuations and uncertainties inherent in their environments.

To address these complexities, we present a multi-agent Retrieval-Augmented Generation (RAG) system designed to support nuclear waste management governance by enhancing regulatory compliance assessments and risk evaluation. This system leverages large language models (LLMs) integrated with document retrieval mechanisms to ensure that agent-generated responses remain grounded in authoritative regulatory and safety guidelines. Structured around multi-agent collaboration, the system incorporates specialized agents such as Regulatory Compliance, Safety & Environmental, and Documentation & Reporting Agents that iteratively refine their analyses through structured discussions, reducing semantic drift and improving decision accuracy. The system operates on consumer-grade hardware using Llama 3.2 [9] and mxbai-embed-large-v1 [10] for embedding generation, ensuring accessibility while maintaining performance. Through structured 10-round discussions, agents collaboratively assess regulatory compliance and safety requirements, grounding their decisions in authoritative documentation through RAG-enhanced information retrieval. We demonstrate the system's effectiveness through a case study of a proposed temporary nuclear waste storage site near Winslow, Arizona. We evaluate performance through metrics including context relevance distribution and semantic drift analysis. This research advances the field by providing an empirically validated framework for AI-assisted regulatory decision-making in complex environmental governance scenarios.

## SYSTEM OVERVIEW

### Principal Decision-making Phases

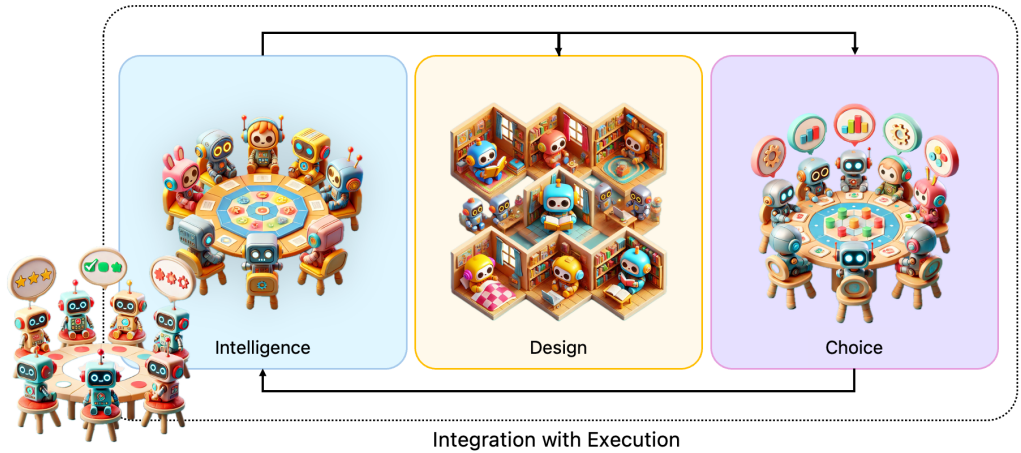

Figure 1. Simon's Three Principal Decision-making Phases [11], [12].

Herbert A. Simon's decision-making model, as described in Figure 1, delineates three interrelated phases—Intelligence, Design, and Choice—each vital to effective problem-solving [11], [12]. In the Intelligence Phase, decision-makers scrutinize their environment to identify problems or opportunities, engaging in activities such as assessing economic, technical, political, and social factors, defining the issues at hand, and collecting pertinent information. The Design Phase focuses on crafting and evaluating potential solutions, fostering innovation, and addressing sub-problems, often necessitating a return to earlier phases for refinement. During the Choice Phase, the optimal alternative is selected by carefully assessing trade-offs, risks, and alignment with overarching goals, typically culminating in implementation and subsequent decision requirements. Eigner and Händler's framework [13] excelled in their framework concerning theoretical concepts and single-agent systems, as it examines decision-making processes where a user interacts with a single LLM for assistance.

We focus on multi-agent systems and aim to provide empirical evidence of their practical application with detailed insights. A multi-LLM agent-based framework for nuclear waste management decision-making serves as an example of a structured and interdisciplinary approach to handling this complex process. This framework includes specialized agents aligned with Herbert Simon's phases of Intelligence, Design, and Choice. By integrating diverse expertise, it effectively addresses the technical, regulatory, social, and environmental dimensions of nuclear waste management.

*Intelligence Phase*
During this phase, the framework gathers and synthesizes critical data to define challenges and identify opportunities. The Regulatory Compliance Agent ensures adherence to both national standards and international guidelines, while the Safety & Environmental Agent assesses ecological and safety risks over the long term. The Knowledge Base & Research Agent curates and applies the latest scientific and technical advancements, maintaining a repository of relevant findings to inform decisions. The Public Relations & Communication Agent captures stakeholder perspectives and public sentiment, promoting transparency and addressing societal concerns. These coordinated efforts establish a well-rounded understanding of key issues like site feasibility, transport risks, and public opposition.

*Design Phase*
Collaboration among agents characterizes this stage, where solutions are crafted, evaluated, and refined. The Safety & Environmental Agent ensures environmental and safety standards are upheld, while the Transportation & Logistics Agent focuses on secure nuclear waste transit. The Public Relations &

Communication Agent facilitates dialogue to incorporate community priorities, building trust and improving solution acceptance. Meanwhile, the Knowledge Base & Research Agent supports iterative solution improvements with cutting-edge research. The Monitoring and Surveillance Agent conducts simulations and real-time anomaly detection to enhance reliability, testing solution robustness before implementation. This collaborative approach ensures that proposed strategies are technically feasible, environmentally sustainable, logistically efficient, and socially acceptable.

*Choice Phase*
In this phase, agents collectively prioritize and recommend the optimal course of action. The Regulatory Compliance Agent verifies legal adherence, while the Safety & Environmental Agent evaluates long-term impacts and risk mitigation. The Public Relations & Communication Agent addresses stakeholder feedback to align public and institutional support. The Incident Response Agent anticipates potential emergencies, ensuring readiness for operational challenges. Lastly, the Documentation & Reporting Agent ensures transparency through detailed records of the decision-making process, promoting accountability and clarity. This phase culminates in selecting and preparing to implement the best strategy while planning for execution challenges.

*Iterative and Adaptive Governance*
This framework acknowledges that nuclear waste management involves a cyclical process rather than a linear one, with obstacles often necessitating revisiting earlier phases for adjustment. Decisions are part of an intricate, multi-level system, described as "wheels within wheels, [11, p. 41]" where strategies must remain flexible and adaptive. By leveraging interdisciplinary AI expertise, the multi-LLM framework fosters inclusivity, scenario modeling, and iterative improvements, supporting dynamic governance processes that respond effectively to evolving conditions. For example, selecting a nuclear waste repository site would require the Regulatory Compliance Agent to ensure international standard adherence, the Safety & Environmental Agent to evaluate geological stability and risks, and the Public Relations & Communication Agent to build public trust. This collaborative and transparent methodology ensures that decisions are grounded in expertise while addressing technical and social complexities. Through such an integrated approach, the framework promotes sustainable, well-supported solutions to the challenges of nuclear waste management.

**DESCRIPTION**

**Agents Overview**
Table 1 presents the eight potential agents involved in effective nuclear waste management governance and briefly explains their respective functions. These agents work in a coordinated manner to address the technical, regulatory, social, and environmental complexities of the process. The 1) **Regulatory Compliance Agent** ensures that all actions adhere to national and international regulations, providing crucial updates to other agents to maintain compliance. The 2) **Safety & Environmental Agent** performs safety evaluations and environmental impact assessments, sharing its findings with monitoring, logistics, and communication agents to inform risk-based decision-making. The 3) **Knowledge Base & Research Agent** integrates research and development with ongoing knowledge maintenance, delivering critical insights to the safety, compliance, and emergency agents to support evidence-based actions. The 4) **Monitoring and Surveillance Agent** plays a vital role in real-time operational oversight by identifying anomalies and alerting safety, logistics, and emergency response agents to address emerging issues promptly. The 5) **Public Relations & Communication Agent** ensures transparency and stakeholder engagement by coordinating with safety, compliance, and emergency response agents to communicate risks and strategies effectively. The 6) **Transportation & Logistics Agent** manages the secure movement of nuclear materials, working alongside compliance, monitoring, and emergency response agents to maintain safety and regulatory standards. The 7) **Incident Response Agent** handles emergencies by collaborating with monitoring, communication, and compliance agents to ensure swift, coordinated responses. Finally, the 8) **Documentation & Reporting Agent** collects and compiles data from all agents, producing detailed reports that enhance accountability, traceability, and transparency. Together, these agents form a cohesive and adaptive framework to address the intricate challenges of nuclear waste governance.

Table 1. Agents and Their Functions in Nuclear Waste Management Governance.

| # | Agents | Functions |
|---|--------|-----------|
| 1 | Regulatory Compliance Agent | Combines national oversight and international regulations. |
| 2 | Safety & Environmental Agent | Covers both safety and environmental impact assessments. |
| 3 | Knowledge Base & Research Agent | Merges R&D and knowledge maintenance. |
| 4 | Monitoring and Surveillance Agent | Handles real-time monitoring and anomaly detection. |
| 5 | Public Relations & Communication Agent | Integrates stakeholder communication and public relations. |
| 6 | Transportation & Logistics Agent | Manages logistics for nuclear waste transportation. |
| 7 | Incident Response Agent | Handles emergency and incident response. |
| 8 | Documentation & Reporting Agent | Manages reporting and documentation processes. |

*Base Large Language Models*
We recognize that using cloud-based large language models (LLMs) is not sustainable, both financially and environmentally. Therefore, we aim to develop a system that operates effectively on consumer-level computers, even those without a GPU or with limited GPU capabilities. This approach is crucial for supporting decision-makers and policy practitioners. To achieve this, we have chosen to base our model on Llama 3.2 created by Meta [9], which has demonstrated outstanding performance with a minimal number of parameters. For embeddings—representations of the semantic meanings of words, phrases, or concepts in numerical vector format—we are utilizing mxbai-embed-large-v1 created by Mixedbread [10], which outperforms closed commercial models like OpenAI [14]'s text-embedding-v3. These embeddings allow our LLM to understand context and relationships, enabling it to process and analyze inputs effectively, and provide relevant and context-aware responses. All model operations are conducted via Ollama [15].

Retrieval-Augmented Generation (RAG) combines retrieval systems with Large Language Models (LLMs) to enhance response generation through external knowledge integration. This hybrid architecture, discussed extensively in recent literature, leverages embedding technology to transform data into high-dimensional numerical vectors where semantic proximity indicates meaning similarity. For instance, vectors representing "dog food" and "canine nutrition" would cluster closely in the embedding space, demonstrating semantic relationships beyond simple keyword matching. The embedding-based similarity search mechanism in RAG retrieves contextually relevant information from knowledge bases to ground the generative model's output in factual content. This integration of embeddings with RAG creates a sophisticated system that produces responses anchored in external knowledge while maintaining semantic understanding. The synergistic relationship between embeddings and RAG has revolutionized information processing by enabling AI systems to generate more accurate, contextual, and informative content through the seamless fusion of retrieval and generation capabilities.

**Evaluation Metrics**

In assessing the performance of multi-agent systems for regulatory compliance, a comprehensive evaluation framework employs four distinct metrics to measure various aspects of system effectiveness. These metrics address both the quality of document-based reasoning and the dynamics of agent interactions, providing quantitative insights into system performance. The evaluation framework encompasses Relevance Score Distribution to measure document alignment, Agent Agreement Rate to assess decision consistency, Conversation-Document Mapping to track information grounding, and Semantic Drift Detection to monitor logical coherence. Together, these metrics offer a multifaceted approach to understanding and validating the system's capability in handling complex regulatory assessments. Each metric serves a specific analytical purpose while complementing the others to provide a holistic view of system performance.

*Relevance Score Distribution*
Relevance scores evaluate the similarity between an agent's responses and the document content, using metrics like Cosine Similarity. This score helps quantify how well the agent's responses align with the document context. Cosine Similarity calculates the similarity between two vectors $(A, B)$:, $where\ A \cdot B$ is the dot product of the vectors, and $|A|, |B|$ are their magnitudes. Relevance scores are visualized as a histogram with relevance scores (0–1) on the X-axis and the frequency of responses on the Y-axis. This metric provides insights into the overall distribution of response relevance.

*Agent Agreement Rate*
The Agent Agreement Rate evaluates the alignment between agents, such as a Safety Agent and a Regulator Agent, when performing the same task. Agreement Rate is calculated as $Agreement\ Rate\ = \frac{Number\ of\ Agree\ Decisions}{Total\ number\ of\ Decisions}$. Visualization is performed using a pie chart to depict the proportion of "Agree" and "Disagree" outcomes. This metric highlights the consistency in decision-making across

*Conversation-Document Mapping*
Conversation-document mapping analyzes how an agent's responses correspond to specific sections of the document, ensuring the conversation remains grounded in the document's content. The structure includes nodes representing document sections and relevant hashtags and edges indicating connections between agents and the sections they reference. For example, a Safety Agent might reference "Paragraph 1" in "Document A," while a Regulator Agent cites "Paragraph 2" in "Document B." This mapping is visualized in graph form, revealing the relationships between the document and conversational elements, thus assessing how effectively agents utilize the provided documentation.

*Semantic Drift Detection*
Semantic drift detection identifies deviations in the conversation's logical context, ensuring that responses remain relevant to preceding questions and the document content. The Semantic Similarity Matrix measures the semantic consistency between questions and responses, where $Q_i$ and $R_j$ are vector representations of question $i$ and response $j$. Heatmaps display the similarity matrix, with low similarity values highlighting potential semantic drift. This metric ensures logical coherence in the flow of conversation and helps detect potential misalignments.

**CASE STUDY: NUCLEAR WASTE STORAGE SITE**

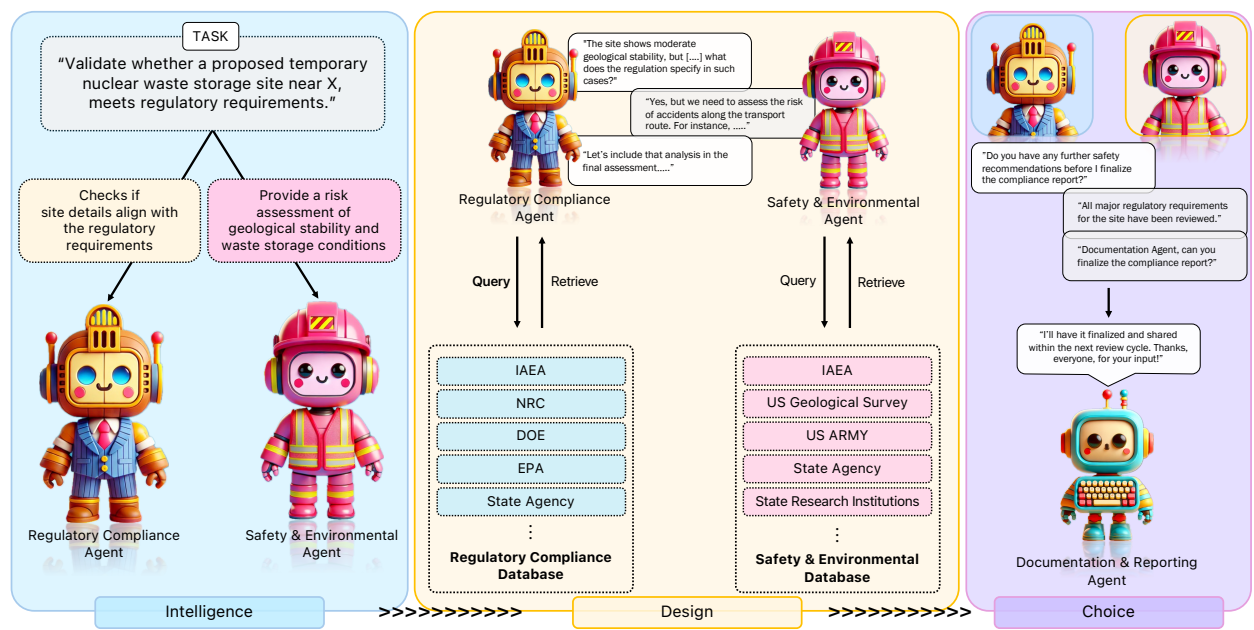

Figure 2. Multi-Agent System for Regulatory Compliance and Safety Assessment in Nuclear Waste Storage Validation.

Figure 2 illustrates a multi-agent system designed to validate whether a proposed temporary nuclear waste storage site near Winslow, Arizona, meets basic national regulatory requirements. This system operates using RAG integrating regulatory compliance and safety assessments through structured agent discussions, document retrieval, and iterative validation.

**System Structure**
The system comprises three key agents: the Regulatory Compliance Agent, the Safety & Environmental Agent, and the Documentation & Reporting Agent. These agents collaborate to assess the proposed site based on legal, environmental, and safety regulations, retrieving authoritative documents and engaging in structured dialogue.

1. **Regulatory Compliance Agent (RCA)** ensures that the site meets national regulatory requirements. It queries regulatory databases such as the International Atomic Energy Agency (IAEA), the Nuclear Regulatory Commission (NRC), the Department of Energy (DOE), the Environmental Protection Agency (EPA), and state agencies. RCA verifies compliance by evaluating laws and quality assurance guidelines for nuclear waste storage.
2. **Safety & Environmental Agent (SEA)** assesses geological stability, environmental risks, and long-term safety measures. It queries sources such as the U.S. Geological Survey, the U.S. Army, state research institutions, and the IAEA, retrieving data on geological conditions, seismic activity, and transport risks.
3. **Documentation & Reporting Agent (DRA)** compiles findings from RCA and SEA into a final compliance report. This agent ensures that all regulatory and safety concerns are addressed before the site assessment is finalized.

Each agent retrieves relevant information, discusses findings in a structured dialogue, and contributes to a final decision, ensuring a comprehensive regulatory validation process.

**Query Processing & Agent Dialogue**

The system processes queries dynamically, following a structured 10-round discussion model, where RCA and SEA refine their assessments iteratively. Queries are rewritten and optimized for precision, recall, and document relevance, leveraging retrieved information to generate contextually grounded responses.

For instance, when given the task:

*"Validate whether a proposed temporary nuclear waste storage site near Winslow, Arizona, meets basic national regulatory requirements."*

- RCA initiates by querying regulatory guidelines, asking: What regulations govern temporary nuclear waste storage?
- SEA queries safety conditions, asking: What are the geological and environmental risks for the Winslow site?
- The discussion evolves, addressing site stability, regulatory gaps, emergency planning, and public safety before finalizing a report.

An example query-response cycle reflects this structured process:
- Query: What specific measures are required to control radiation exposure and the release of radioactive material?
- Retrieved Context: Fundamental Safety Principles (IAEA), Hazardous Waste Storage Regulations (AZDEQ), and Quality Assurance Program Guide (DOE).
- Generated Response: Regulations require restricting exposure through controlled facility design, emergency response planning, and waste containment measures.

This iterative validation ensures responses are grounded in retrieved knowledge rather than static model outputs, improving factual accuracy and regulatory alignment.

**Document Retrieval & Compliance Validation**
The system retrieves information from a predefined document repository, containing key regulatory and safety guidelines. The document structure is hierarchically organized under two main categories:
1. Regulatory Compliance Documents (RCA)
    - DOE Quality Assurance Guides (DOE G 414.1-1C, 414.1-2B, 414.1-4) → Ensure compliance in facility design, waste handling, and operational safety.
    - IAEA Safety Principles & Waste Disposal Regulations → Provide international safety benchmarks.
    - AZDEQ Hazardous Waste Regulations → Establish state-level compliance requirements.

2. Safety & Environmental Documents (SEA)
    - IAEA Safety Analysis Reports & Geological Disposal Facilities → Assess environmental and geological risks.
    - U.S. Geological Survey & U.S. Army Reports → Provide geotechnical data relevant to site stability.

When agents query the system, document retrieval prioritizes relevance, extracting contextual passages to support decision-making. For example:
- If RCA queries, "What are the regulatory requirements for nuclear waste storage?" it retrieves DOE, NRC, and IAEA regulations.
- If SEA queries, "How do geological risks impact the Winslow site?" it pulls data from U.S. Geological Survey reports.

This retrieval-augmented approach ensures agent discussions are grounded in factual, up-to-date regulatory and safety standards, minimizing semantic drift and misinformation.

**Final Decision & Reporting**
After RCA and SEA complete their structured discussions, DRA compiles a compliance report summarizing key findings:
- Regulatory Status → Does the Winslow site meet national and state requirements?
- Environmental & Safety Assessment → Are there geological risks or transport safety concerns?
- Mitigation & Emergency Plans → How does the site address long-term safety?

If all requirements are met, the site receives preliminary approval. If gaps remain, further risk assessments or policy modifications may be necessary. This structured multi-agent validation process ensures that nuclear waste storage sites undergo rigorous, document-backed compliance evaluation.

**RESULTS**

**Context Relevance Distribution**

The context relevance distribution, as depicted in Figure 4, analyzes document-grounded response generation effectiveness between the Safety Agent and Regulatory Compliance Agent. Cosine Similarity scores measure alignment between agent responses and authoritative sources, indicating factual consistency in decision-making. Analysis reveals the Regulatory Agent achieves higher median relevance scores across topics, demonstrating stronger alignment with legal frameworks due to structured regulatory documentation.

The Safety Agent exhibits wider score variability, particularly in Topic 1, reflecting the inherent complexity of environmental and safety assessments that demand broader contextual reasoning. This variation stems from the need to integrate diverse data sources and consider multiple risk factors simultaneously. Despite these differences, both agents maintain robust document grounding throughout their assessments. The Regulatory Agent excels in precise legal interpretations, while the Safety Agent effectively manages complex risk assessments requiring multifaceted analysis. These complementary strengths validate the multi-agent RAG framework's ability to handle diverse aspects of regulatory compliance through structured discussions. The distribution patterns confirm the system's capacity to maintain document relevance while addressing varying complexity levels in nuclear waste management governance.

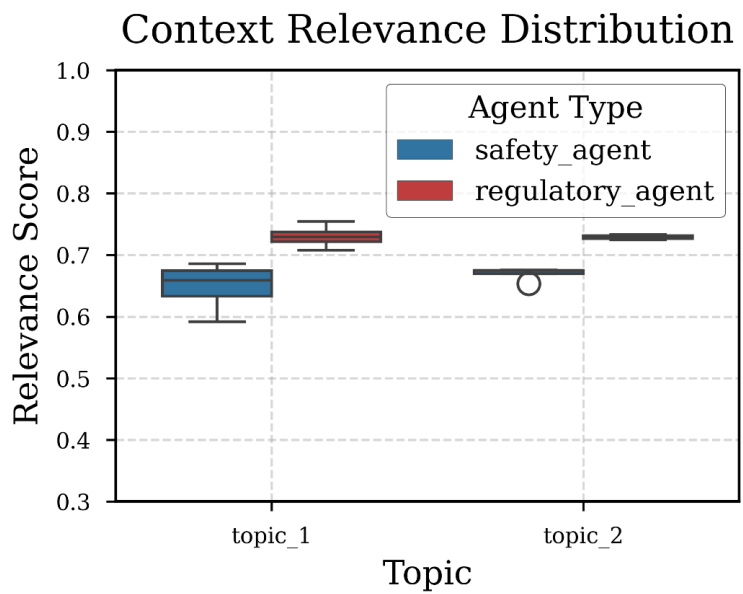

Figure 3. Context Relevance Distribution for Safety and Regulatory Agents.
*This figure represents the relevance score distribution for two agent types: Safety Agent (blue) and Regulatory Agent (red), evaluated across two topics (Topic 1 and Topic 2). The Y-axis represents the relevance score, ranging from 0.3 to 1.0, quantifying agent response alignment with retrieved document content. The box plots display the median, interquartile range, and variability in relevance scores. Higher scores indicate a stronger grounding of responses in regulatory and safety documentation, while the spread of the distribution reflects variations in response consistency.*

**Agreement Rate and Semantic Drift Analysis**
Figure 5 illustrates the Agreement Rate and Semantic Drift Analysis, measuring consistency in multi-agent discussions through the RAG framework. Agreement Rate tracks decision-making alignment between the Regulatory Compliance Agent and Safety & Environmental Agent, while Semantic Drift quantifies response deviations from document grounding. Agreement Rates show initial fluctuations, particularly in Topic 2's early rounds, indicating iterative response refinement before reaching a consensus. Simultaneously, decreasing Semantic Drift trends demonstrate improved response coherence and reduced inconsistencies. The inverse relationship between rising Agreement Rates and declining Semantic Drift confirms effective regulatory assessment refinement through structured discussions. These trends validate how the multi-agent framework enhances decision-making consistency across successive interactions. Agents progressively strengthen alignment with regulatory and safety documentation while minimizing response variability, ensuring robust compliance validation. The complementary metrics demonstrate the system's ability to maintain factual grounding while fostering inter-agent consensus, which is essential for reliable nuclear waste management governance.

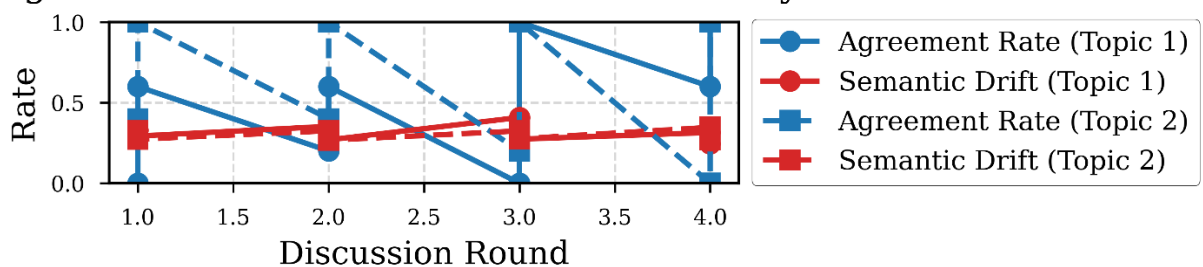

Figure 4. Agreement Rate and Semantic Drift Across Discussion Rounds.
*This figure presents the Agent Agreement Rate (blue) and Semantic Drift (red) across four discussion rounds for Topic 1 (solid lines with circular markers) and Topic 2 (dashed lines with square markers). The X-axis represents the discussion rounds, while the Y-axis measures the rate of agreement and semantic drift. Higher agreement rates indicate increasing consensus between agents, whereas declining semantic drift suggests improved conversational coherence and alignment with retrieved documents. The dashed lines connecting data points serve to visualize trends and do not represent continuous measurements between discussion rounds.*

**DISCUSSION**

*Interpretation of Results*
The results presented in the previous sections demonstrate the effectiveness of the multi-agent discussion framework in ensuring document-grounded responses for nuclear waste site compliance assessments. The progressive improvements in context relevance distribution and agreement rate trends indicate that the system successfully refines its response accuracy and coherence across iterative discussions. This section examines these findings in greater detail, comparing them with expected outcomes, analyzing performance trends, discussing potential limitations, and identifying future research directions.

The context relevance distribution demonstrates that both the Regulatory Compliance Agent (RCA) and the Safety & Environmental Agent (SEA) generate responses with strong alignment to the retrieved documents. While both agents exhibited comparable relevance distributions, the Regulatory Agent consistently achieved slightly higher median and upper-quartile relevance scores, indicating greater precision in regulatory interpretations. This finding suggests that regulatory constraints, typically well-defined in structured guidelines, facilitate more precise information alignment compared to safety-related risk assessments, which may involve more complex, context-dependent decision-making.

The Agreement Rate and Semantic Drift Analysis further validates the system's effectiveness by showing an increase in agreement rates and a corresponding decrease in semantic drift over successive discussion rounds. As agents refine their responses through structured dialogues, they reach higher levels of consensus, ensuring decision-making consistency in nuclear waste governance. The inverse relationship between agreement and semantic drift is particularly significant, as it highlights the system's ability to maintain response coherence while integrating new information across iterative retrieval cycles.

The multi-agent framework demonstrates greater adaptability and contextual awareness than traditional rule-based regulatory compliance systems. The structured 10-round discussion model used in this system provides a novel approach to incremental knowledge refinement, improving both decision coherence and response accuracy. Our implementation demonstrates distinct advantages over traditional single-agent systems, particularly in handling complex regulatory documentation. The system's ability to maintain high document relevance while supporting inter-agent dialogue represents a significant advancement in automated regulatory compliance assessment. However, we acknowledge potential limitations in processing ambiguous regulatory language, where human expertise remains crucial for final decision-making.

**Potential Limitations**
Despite its strengths, the system presents several potential limitations. First, retrieval effectiveness remains contingent on the quality and availability of external documents. If a relevant regulatory update is missing from the knowledge base, the system may generate partially incomplete responses, impacting compliance accuracy. Second, while effective for accuracy, the discussion-based refinement approach introduces additional computational overhead, which may impact real-time decision-making in high-stakes operational environments, such as incident response scenarios. Third, although the system minimizes semantic drift, it does not entirely eliminate the possibility of minor response inconsistencies, particularly when integrating multiple regulatory interpretations. Fourth, the RAG metrics reveal challenges in maintaining consistent retrieval performance across different document types and topics. The system's Precision, Recall, and F1-scores can fluctuate depending on document complexity and query specificity, potentially affecting the completeness of regulatory assessments. Additionally, the embedding-based retrieval mechanism may occasionally struggle with nuanced regulatory language, leading to suboptimal document chunk selection. Future enhancements should explore reinforcement learning techniques to optimize response coherence in complex decision-making tasks.

**Implications and Future Research**
The findings suggest that multi-agent, retrieval-augmented systems have significant potential for improving regulatory compliance assessments in nuclear waste governance. Future research should explore enhancing agent specialization, where distinct agents are fine-tuned to handle specific regulatory domains, such as transportation safety, long-term waste storage, and public health risk assessments. Additionally, incorporating more sophisticated reasoning mechanisms, such as causal inference models, could further enhance the system's ability to evaluate regulatory trade-offs in real-world scenarios.

Another important direction involves scalability testing, particularly in decentralized deployment environments where decision-making must be distributed across multiple governmental and private-sector stakeholders. Ensuring that the system remains computationally efficient while maintaining high accuracy is crucial for real-world implementation. Future iterations could also explore integrating human-in-the-loop verification mechanisms, allowing domain experts to provide feedback on agent-generated responses, further improving regulatory compliance validation processes.

Moreover, expanding the system to a multi-modal model could significantly enhance its information processing capabilities. By integrating text, images, satellite data, and sensor readings, agents could more effectively analyze real-time site conditions, geological formations, and structural integrity of waste storage facilities. For instance, incorporating geospatial mapping data alongside retrieved regulatory documents would allow the Safety & Environmental Agent (SEA) to conduct more precise risk assessments regarding geological stability, seismic activity, and potential environmental hazards. This multi-modal capability would further strengthen the system's ability to generate holistic, data-driven compliance evaluations.

Additionally, real-time web search integration could address one of the current system's limitations: reliance on static, predefined document repositories. While the existing retrieval mechanism effectively retrieves information from structured regulatory databases, policy changes, new safety guidelines, and emerging risks require access to continuously updated sources. The system could dynamically retrieve the latest governmental reports, scientific publications, and regulatory amendments by implementing real-time web search capabilities, ensuring agents operate with the most current and relevant information. This enhancement would be particularly beneficial for incident response scenarios, where rapidly evolving situations necessitate up-to-date hazard assessments, emergency protocols, and regulatory advisories.

## CONCLUSIONS

This research presents a transformative multi-agent Retrieval-Augmented Generation (RAG) system for nuclear waste management governance, advancing automated regulatory compliance assessment through structured decision-making. The system integrates document retrieval, semantic embeddings, and structured agent dialogues to generate accurate, contextually relevant responses grounded in regulatory guidelines and safety assessments. The Regulatory Compliance Agent (RCA) exhibits superior precision in legal interpretations, ensuring strict adherence to national and international regulatory frameworks. At the same time, the Safety & Environmental Agent (SEA) effectively manages complex risk assessments, addressing long-term site stability and ecological considerations. The inverse correlation between increasing agreement rates and decreasing semantic drift further validates the system's ability to enhance consistency in agent interactions, ensuring that responses remain factually accurate and aligned with evolving regulatory conditions.

The framework surpasses traditional rule-based regulatory systems, which rely on static predefined queries, by enabling dynamic response adjustment to evolving regulatory conditions. Integrating real-time document retrieval with structured agent dialogue supports a more adaptive and scalable decision-support framework, particularly valuable in complex, interdisciplinary contexts where multiple regulatory frameworks intersect. The retrieval-augmented approach ensures that agents operate with domain-specific knowledge, improving decision-making relevance across diverse scenarios, such as nuclear waste transportation, long-term repository siting, and emergency response preparedness.

Several limitations merit consideration. System performance depends on document availability, meaning that missing or outdated regulations could introduce gaps in compliance validation. Additionally, real-time decision-making demands significant computational resources, particularly in emergency response scenarios, where agents must process high volumes of real-time data while maintaining response accuracy. Although semantic drift minimization proves effective, minor response inconsistencies may emerge when reconciling multiple regulatory interpretations due to varying levels of specificity in legal frameworks. These challenges underscore the importance of continued system refinement and validation, particularly for high-stakes applications requiring precise regulatory interpretations and dynamic risk assessments.

Future research directions include reinforcement learning to enhance response coherence, allowing the system to optimize regulatory assessments based on historical decision trends. Additionally, human-in-the-loop verification mechanisms could integrate expert oversight, ensuring that AI-generated responses remain legally and ethically sound in real-world applications. Expanding the system to multi-modal processing capabilities would further improve decision-making accuracy, particularly in areas where text-based analysis alone is insufficient. For instance, integrating satellite imagery, sensor data, and geospatial mapping would enable agents to assess site stability, detect environmental anomalies, and validate compliance with geological safety requirements in real time. Furthermore, incorporating dynamic web retrieval of policy updates and scientific publications would ensure that agent operations remain informed by the latest regulatory developments and emerging risk factors, a crucial capability for rapidly evolving emergency situations where updated guidance is essential for effective decision-making.

This framework establishes a scalable, adaptable approach for addressing complex governance challenges in nuclear waste management. The system advances AI-driven compliance validation by combining RAG technology, structured agent interactions, and iterative validation mechanisms while ensuring accountability, transparency, and regulatory robustness in high-stakes environmental management scenarios. The findings demonstrate significant potential for improving regulatory decision-making through AI systems that balance automation with human oversight, fostering a responsible and resilient governance framework for nuclear waste storage, transportation, and long-term safety planning.

## ACKNOWLEDGMENTS

We sincerely thank the anonymous reviewers for their insightful comments and constructive suggestions that significantly improved the quality of this manuscript. The computations in this paper were conducted on the Sol supercomputer at Arizona State University [16]. This project was partially supported by the Arizona State University Graduate and Professional Student Association's JumpStart Grant Program. Any opinions, findings, and conclusions expressed in this material are those of the authors and do not necessarily reflect the views of the funding agencies.